\documentstyle{mn}

\input{epsf}

\newif\ifAMStwofonts

\ifoldfss
  \ifCUPmtlplainloaded \else
    \NewTextAlphabet{textbfit} {cmbxti10} {}
    \NewTextAlphabet{textbfss} {cmssbx10} {}
    \NewMathAlphabet{mathbfit} {cmbxti10} {} 
    \NewMathAlphabet{mathbfss} {cmssbx10} {} 
  \fi
  \ifAMStwofonts
    \ifCUPmtlplainloaded \else
      \NewSymbolFont{upmath} {eurm10}
      \NewSymbolFont{AMSa} {msam10}
      \NewMathSymbol{\upi}     {0}{upmath}{19}
      \NewMathSymbol{\umu}     {0}{upmath}{16}
      \NewMathSymbol{\upartial}{0}{upmath}{40}
      \NewMathSymbol{\leqslant}{3}{AMSa}{36}
      \NewMathSymbol{\geqslant}{3}{AMSa}{3E}

       \let\le=\leqslant
       
    \fi
  \fi
\fi 

\ifnfssone
  \newmathalphabet{\mathit}
  \addtoversion{normal}{\mathit}{cmr}{m}{it}
  \addtoversion{bold}{\mathit}{cmr}{bx}{it}
  \newmathalphabet{\mathbfit} 
  \addtoversion{normal}{\mathbfit}{cmr}{bx}{it}
  \addtoversion{bold}{\mathbfit}{cmr}{bx}{it}
  \newmathalphabet{\mathbfss} 
  \addtoversion{normal}{\mathbfss}{cmss}{bx}{n}
  \addtoversion{bold}{\mathbfss}{cmss}{bx}{n}
  \ifAMStwofonts
    \ifCUPmtlplainloaded \else
      \UseAMStwoboldmath
      \makeatletter
      \new@mathgroup\upmath@group
      \define@mathgroup\mv@normal\upmath@group{eur}{m}{n}
      \define@mathgroup\mv@bold\upmath@group{eur}{b}{n}
      \edef\UPM{\hexnumber\upmath@group}
      \new@mathgroup\amsa@group
      \define@mathgroup\mv@normal\amsa@group{msa}{m}{n}
      \define@mathgroup\mv@bold\amsa@group{msa}{m}{n}
      \edef\AMSa{\hexnumber\amsa@group}
      \makeatother
      \mathchardef\upi="0\UPM19
      \mathchardef\umu="0\UPM16
      \mathchardef\upartial="0\UPM40
      \mathchardef\leqslant="3\AMSa36
      \mathchardef\geqslant="3\AMSa3E

       \let\le=\leqslant

    \fi
  \fi
\fi 

\ifnfsstwo
  \DeclareMathAlphabet{\mathbfit}{OT1}{cmr}{bx}{it}
  \SetMathAlphabet\mathbfit{bold}{OT1}{cmr}{bx}{it}
  \DeclareMathAlphabet{\mathbfss}{OT1}{cmss}{bx}{n}
  \SetMathAlphabet\mathbfss{bold}{OT1}{cmss}{bx}{n}
  \ifAMStwofonts
    \ifCUPmtlplainloaded \else
      \DeclareSymbolFont{UPM}{U}{eur}{m}{n}
      \SetSymbolFont{UPM}{bold}{U}{eur}{b}{n}
      \DeclareSymbolFont{AMSa}{U}{msa}{m}{n}
      \DeclareMathSymbol{\upi}{0}{UPM}{"19}
      \DeclareMathSymbol{\umu}{0}{UPM}{"16}
      \DeclareMathSymbol{\upartial}{0}{UPM}{"40}
      \DeclareMathSymbol{\leqslant}{3}{AMSa}{"36}
      \DeclareMathSymbol{\geqslant}{3}{AMSa}{"3E}

       \let\le=\leqslant

    \fi
  \fi
\fi 

\ifCUPmtlplainloaded \else
  \ifAMStwofonts \else 
    \def\upi{\pi}
    \def\umu{\mu}
    \def\upartial{\partial}
  \fi
\fi

\def\sigmah{\sigma_{\rm H}}

\title{The signature of dark energy on the local Hubble flow}
\author[A.V. Macci\`o et al.]
{Andrea V. Macci\`o$^1$ \thanks{E-mail: andrea@physik.unizh.ch}, Fabio
  Governato$^{2,3}$ \& Cathy Horellou$^4$ \\
$^1$ Institute for Theoretical Physics, University of Z$\ddot u$rich, CH-8057
  Z$\ddot u$rich, Switzerland \\
$^2$ Department of Astronomy, University of Washington, Seattle, WA 98195, USA\\  
$^3$ INAF--Osservatorio Astronomico di Brera, Milano, Italy \\ 
$^4$ Onsala Space Observatory, Chalmers University of Technology, SE-43992 Onsala, Sweden\\ 
}
\smallskip 

\pubyear{2005}

\begin{document}

\maketitle


\begin{abstract}

Using $N$-body simulations of flat, dark energy dominated cosmologies,
we show that galaxies around simulated binary systems resembling the
Local Group (LG) have low peculiar velocities, in good agreement with
observational data.  We have compared results for LG-like systems
selected from large, high resolution simulations of three cosmologies:
a $\Lambda$CDM model, a $\Lambda$WDM model with a 2~keV warm dark
matter candidate and a quintessence model (QCDM) with an equation of
state parameter $w=-0.6$.

The Hubble flow is significant colder around LGs selected in a flat,
$\Lambda$ dominated cosmology than around LGs in open or critical
models, showing that a dark energy component manifests itself on the
scales of nearby galaxies, cooling galaxy peculiar motions.  Flows in
the $\Lambda$WDM and QCDM models are marginally colder than in the
$\Lambda$CDM one.

The results of our simulations have been compared to existing data and
a new data set of 28 nearby galaxies with robust distance measures
(Cepheids and Surface Brightness Fluctuations).  The measured
line-of-sight velocity dispersion is $\sigmah$ = 88$\pm$ 20 {\rm
km/sec} $\times$ (R/7 {\rm Mpc}).  The best agreement with
observations is found for LGs selected in the $\Lambda$CDM cosmology
in environments with $-0.1 < \delta \rho / \rho < 0.6$ on scales of
7~Mpc, in agreement with existing observational estimates on the local
matter density.

These results provide new, independent evidence for the presence of
dark energy on scales of few Mpc, corroborating the evidence gathered
from observations of distant objects and the early Universe.

\end{abstract}

\begin{keywords}
cosmology: theory -- dark matter -- large-scale structure of
the Universe -- galaxies: clusters -- galaxies: haloes -- methods: numerical
\end{keywords}

\newpage

\section{Introduction}

The Hubble flow just outside the Local Group is {\em dynamically cold}, 
as indicated by 
the small scatter in the velocity-distance diagram of nearby galaxies.  
This was first pointed out by 
Sandage, Tammann \& Hardy (1972). 
The peculiar line-of-sight velocity dispersion $\sigmah$
(see eq.~\ref{eq:HF} for a definition) of galaxies within $5-7 h^{-1}$ Mpc
lies in the range $80 - 40$~km~s$^{-1}$
(Giraud 1986, Schlegel et al. 1994, Sandage 1999, Ekholm et al. 2001, 
Karachentsev et al. 2003b). 
This low value is likely linked to the properties of the energy-matter density
field around the Local Group and it can be used to constrain cosmological models.

Interestingly,  cosmological  
cold dark matter (CDM) simulations of hierarchical structure formation 
have so far not been able 
to produce values of the velocity dispersion around simulated galaxy groups 
as low as the observed one. 
The three models tested in the works of 
Schlegel et al. (1994) and Governato et al. (1997, hereafter G97) all showed substantially
larger velocity dispersions and failed to recover a significant number
of binary systems with the properties of our Local Group. 
Typical values of $\sigmah$ ranged from 200~km/sec (for the open model) to  
500~km/sec (for the old ``standard" CDM model with $\Omega=1$), 
several times higher than the observational estimates.

Since these early works, substantial progress has been made in the determination of the
main cosmological parameters. Observations indicate  
the presence of a significant component of smooth
energy with large negative pressure 
(observations of high-redshift supernovae by Perlmutter et al. 1999, Riess et al. 1999;
analysis of fluctuations of the cosmic microwave background combined with data on the
large-scale structure of galactic distribution by 
Balbi et al. 2000, Tegmark,
Zaldarriaga \& Hamilton 2001, Netterfield et al. 2002, Spergel et al. 2003,
Verde et al. 2003, Dai, Liang \& Xu 2004). 
This component has been dubbed dark energy, and
the most appealing candidate is Einstein's 
cosmological constant $\Lambda$ (see Peebles \& Ratra 2003 for a recent review).

On theoretical grounds, Baryshev, Chernin \& Teerikorpi (2001)
(but see also Chernin 2001) suggested that a dark energy (DE) component would cause peculiar
velocities to cool adiabatically in regions of the universe where dark
energy overcomes the gravity of the local dark matter concentration. 
In such regions, the usually growing mode of the density
perturbation is decaying ($\delta_g \propto a^{(1+3w)}$ and $w
< -1/3$). 
When dark energy dominates, new structures do not condensate and
linear perturbations of density and peculiar velocities field decay
(this effect was also considered for vacuum-dominated regions by
Chernin et al. 2003). They have shown that for a Milky-Way like halo the
DE starts to be dominant on a scale of few Mpc and that the strength
of this effect is related to the equation of state of the DE.  Their
idea was developed further by Karachentsev, Chernin \& Teerikorpi
(2003) and Chernin et al. (2004), using a numerical approach to trace the
trajectories of the radially expanding galaxies back in time in
different cosmological models to see the effect of a cosmological
constant on the evolution  of the flow.
In a more general frame the importance of DE on Cluster and Galaxy scales was also
pointed out by Mota \& van de Bruck (2004) and Nunes \& Mota (2004) using
linear theory and spherical collapse model and extended to model beyond the
standard FRW model by Teerikorpi, Gromov, Baryshev (2003).

Recently, Klypin et al. (2003a) simulated the evolution of a cosmic region
resembling the real Universe around 100~Mpc from the Local Group.
The density field was mapped using the MARK{\sc iii} survey (Willick et
al. 1997) and a $\Lambda$CDM power spectrum was assumed. The authors report a
peculiar line-of-sight velocity dispersion of $\sim$ 60~km/sec
around the Local Group in their simulation, which is a significant
improvement compared to previous studies. 
We have extended their work on a bigger sample of simulated LGs
to test if the low value of $\sigmah$ reflects 
the specific geometry of the mass distribution, or/and 
is a general characteristic of the cosmology used. 

The possibility that the observational estimate of $\sigmah$ adopted 
in previous works be incorrect or biased 
cannot be dismissed entirely, as
the $\sigma_H$ based on Tully-Fisher (TF) distances 
suffers  likely from large
distance errors (which should however increase $\sigmah$) while the lower value
estimated by Ekholm et al. (2001) was based on a sample of only 
9 galaxies.
Moreover, the galaxies in the TF sample used by Schlegel et al. (1994) 
had been ``regrouped''. 
This was necessary in order to avoid inflating $\sigmah$ with the internal
velocity dispersion of small virialised galaxy associations. 
No detail, however, was given on how the procedure was carried out. 
If, for example, the velocities of 
loose, non virialised associations of galaxies
are averaged, the $\sigmah$ could be effectively ``cooled'' by removing
the peculiar motion of infalling galaxies.

All considered, the existing literature raises interesting questions but 
leaves
them partially unanswered: is the low value of $\sigmah$ a generic problem of
all CDM models or does it depend on the cosmology adopted? How 
important is the influence of the local environment and/or the surrounding 
large scale structure? Is the addition of a cosmological constant important?
How robust is the observational estimate of $\sigmah$ around our own LG?

Motivated by the growing success of flat $\Lambda$-dominated models and the
availability of more 
accurate distance estimates,   
we embarked on a new analysis of the problem. 
In this paper, we present a new sample of distances and velocities of nearby
galaxies combined with results from other authors on the very local ($< 3$ Mpc)
Hubble flow.
We compare these observational data with new
high resolution simulations, considering 
DE models with a constant equation of state 
parameter $w$   different from $-1$ in order  
to test the impact of DE on the cooling 
of the local Hubble flow. 

\section{Revisiting the Coldness of the Hubble Flow}

To obtain a robust estimate of the coldness of the Hubble flow around the Local
Group, it is essential to use accurate distances to nearby galaxies.
As noted above, previous measures of $\sigmah$ were based
on TF distances, which suffer from fairly large relative errors, especially for
small, poorly resolved galaxies and/or viewed at unfavourable angles. 

To this purpose we have gathered data from different sources: Cepheid-based
distances taken from the Hubble Space Telescope (HST) key project to measure the
Hubble constant (Freedman et al. 2001), distance estimates based on the
Surface Brightness Fluctuation method (SBF, Tonry et al. 2001) and, mainly as a
comparison, TF distances (Tully et al. 1992). For galaxies
common to several samples we have kept the Cepheid-based distances. The final
sample contains 28 galaxies within 10~Mpc from the Sun (11 with Cepheid-based
distances; 17 early type galaxies with distances measured using the 
SBF method: 37 with TF distances). We have converted
heliocentric distances into distances relative to the barycenter of the Local
Group, taken  2/3 of the way on the line between M~31 and the Milky Way
(hereafter denoted as $R_{2/3}$). As for the heliocentric distance to M~31, we used 
the Cepheid derived value of 870~kpc (Ekholm et al. 2001). The observed
heliocentric velocities, taken from the NASA Extragalactic Database, were corrected
to the values  they would have if measured from an observer in our
galaxy at rest relative to the centroid of the Local Group, according
to Yahil et al. (1977).

\begin{figure}
\centering
\epsfxsize=\hsize\epsffile{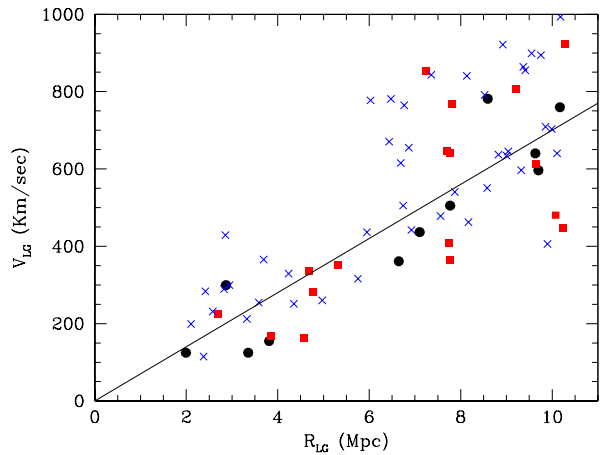}
\caption{
Expansion velocity $V_{LG}$ in the LG reference frame as a function of radial
distance $R_{LG}$ from the Local Group baricenter. Black dots: Cepheid based
distances; Filled squares: Surface Brightness Fluctuations based distances;
Crosses: TF distances. The straight line is for $H_0$=70~km/sec/Mpc.
}
\label{fig:flow}
\end{figure}

With this sample of 28 galaxies 
it is also possible to measure how $\sigmah$ changes with radius as larger and
larger volumes (with radius up to 10~Mpc) are included.
Fig.~\ref{fig:flow} shows a velocity-distance plot, where the distances are relative to the
barycenter of the LG and the velocities have been corrected for the solar
motion  toward the centroid of the Local Group. The crosses show the TF points,
the black dots the Cepheid-based values and the filled squares are from the
SBF sample. The straight line corresponds to $H_0=70$~km~s$^{-1}$~Mpc$^{-1}$.

Measures of $\sigmah$ including the TF sample proved to be
systematically higher by 30--40\% than those using only galaxies with
Cepheid- or SBF-based distances independent of the particular sky
quadrant they were selected, showing that TF galaxies are not tracing
different (hotter) structures, but are rather subject to larger
distance errors. We then decided to not use the TF sample.

We regrouped galaxies in regions with a number overdensity larger than 64. 
This overdensity threshold corresponds roughly to virialised region
(Eke et al. 1996). 
Each group was assigned the mean position and velocity of its members. 
Only a smaller number ($\sim20$\%) of galaxies were linked
together, mostly into binary and triple systems. 
Result obtained using our regrouping scheme, do not differ substantially from those using
galaxy sample without regrouping.

The velocity dispersion increases with the size of the sphere within which
galaxies are included. It varies from 52~km~s$^{-1}$ within 3~Mpc to 135~km~
s$^{-1}$ within 10~Mpc. This is in agreement with previous results of Ekholm
et al. (2001) based on 7 galaxies within 7~Mpc and those  of 
Schlegel et al. (1994), based on a sample of galaxies between 3 and 7~Mpc. 
Our estimate of $\sigmah$ is a bit higher
than the one obtained by Karachentsev et al. (2003b) 
who found $\sigmah \approx 42$~km/sec inside a sphere of 5~Mpc
using distances from
luminosity of the tip of the red giant branch stars of 16 dwarf galaxies, 
but the results are compatible at the $2\sigma$ level.
For $\sigmah$ measured using HST and SBF galaxies in spheres centered on the
LG with growing radius $R$ we 
obtained the following fit: 
$$\sigmah = 88 \pm 20\, {\rm km/sec} \times (R/7\, {\rm Mpc})\, .$$

The final results of the relation $\sigmah$ vs enclosing radius are shown in 
Fig.~\ref{fig:rho}, where we also include the results 
of Karachentsev et al. (2002) for the very local ($D<2.5$~Mpc) Hubble flow.

\section{A Review of theory predictions}
\label{sec:theo}

In this section, we very briefly review the analytical work
done on the subject of cold flows in the context of the Local Group
Hubble flow.  Specifically Baryshev et al. (2001) and Axenides \&
Perivolaropoulos (2002) had already suggested on the basis of simple
dynamical arguments that dark energy could appear as a cooler of
peculiar motions in the Local Group environment.  What is the expected
trend as the constant $w$ in the equation of state of dark
energy changes?

A first fundamental difference between the quintessence  and the
$\Lambda$ ($w=-1$) models is the epoch at which dark energy
dominates over the gravitational attraction of the matter.

The expansion of a flat universe with dark energy is governed by the
Friedmann equation
\begin{equation}
H(a)=\frac{\dot{a}}{a} = H_0 \sqrt{\Omega_0(1+z)^3 + \Omega_{\rm DE,0}(1+z)^{3(1+w)}} \, .
\end{equation}
where $a=1/(1+z)$ is the scale factor. 
The dark energy term dominates over the gravitational attraction of
the matter at redshifts lower than $z_{\rm eq}$: 
\begin{equation}
1+z_{\rm eq} = \left(
\frac{\Omega_{\rm DE,0}}{\Omega_0}
\right)^{-1/3w}\, .
\end{equation}
For $w=-1$, this gives $z_{\rm eq}= 0.326$ and 
for $w=-0.6$ $z_{\rm eq}= 0.601$.

It has been argued from linear perturbation theory that dark energy has
little effect on the local dynamics (Lahav et al. 1991, Wang \& Steinhardt 1998). 
In linear perturbation theory, the peculiar velocity field associated with
a spherical mass concentration with density contrast $\delta$ averaged over
proper radius $R$ is $v_{\rm pec}=\frac{HR}{3} f \delta$, where
$f=\frac{d\ln \delta}{d\ln a}$; $f\simeq \Omega^\alpha$ with 
$\alpha\simeq 0.6$ (Peebles 1980). Wang \& Steinhardt (1998) have provided
an approximation to $\alpha$ in quintessence models, which showed that 
the difference between the linear peculiar velocities in models with different
$w$ is negligible, as the growth factor depends mainly on $\Omega_0$ and dark 
energy dominates only at late times.

However, it has been later recognized that peculiar velocities can be
cooled locally for galaxies located in a ``dark energy dominated
zone'' (Baryshev et al. 2001).  Considering a shell enclosing a mass
$M$ evolving in an expanding background, one can calculate a
``zero-mass" radius beyond which the dark energy dominates over the
gravitational attraction of the matter ($\ddot{r}=0$).  The equation
of motion of the shell is:
\begin{equation}
\ddot{r}= -\frac{GM}{r^2} - \frac{1+3w}{2}\Omega_{\rm DE}(z)H^2(z)r \, .
\end{equation}
Baryshev et al. (2001) have argued that peculiar velocities are cooled
adiabatically for galaxies situated beyond the zero-mass surface, 
where dark energy dominates the dynamics. 
For $w=-1$, the zero-mass radius is independent of the redshift: 
\begin{equation}
r_{\rm zm}^{\Lambda} = \left(
\frac{GM}{\Omega_{\rm DE,0} H_0}
\right)
^{\frac{1}{3}} = 1.36\, {\rm Mpc}
\end{equation}
using $M=2\times10^{12}$~M$_\odot$ for the Local Group 
(van den Bergh 1999) and 
our adopted cosmological parameters. 
For $w\neq-1$, we have 
\begin{equation}
r_{\rm zm}^{\rm DE}= \left(
-\frac{2}{1+3w}
\right)
^{\frac{1}{3}} 
\frac{1}{(1+z)^{(1+w)}}
r_{\rm zm}^\Lambda \, .
\end{equation}

For $w=-1/3$ (which is equivalent to an open model with the same
matter energy content), $r_{\rm zm}\rightarrow +\infty$.
For $w=-0.6$, $r_{\rm zm}^{\rm DE}= \frac{1.357}{(1+z)^{0.4}}r_{\rm
zm}^\Lambda$.
This gives $r_{\rm zm}^{w=-0.6}(z=0)= 1.85$~Mpc.

Note that the relative importance of dark energy on the dynamics of galaxies
around a mass concentration varies in time, depending on the local
density contrast.  The zero-mass radius is larger around high-density
regions, and for a given region it increases with cosmic time.  So a
galaxy which is now close to the Local Group and within the matter
dominated zone could have been in the dark energy dominated zone in
the past (even if the dynamics of the universe as a whole was not dark
energy dominated) and have had its peculiar velocity cooled during a
period of time.  Numerical simulations make it possible to see the
integrated effect of the dynamical evolution of galaxies around
concentrations of mass, which cannot be calculated easily analytically
as galaxies could move in and out of dark energy dominated zones.

If a galaxy has been  in the dark energy dominated zone 
at $r>R_{\rm zm}^{\rm DE}$  
from  redshift $z_{\rm eq}$ until now, 
then its peculiar velocity will have
cooled adiabatically: 
$$v_{\rm pec}(z=0) = \frac{v_{\rm pec}(z_{\rm eq})}{1+z_{\rm eq}}.$$ 
As an upper limit for the peculiar velocity, one can take the Hubble
expansion velocity $v_{\rm H} = HR$: 
$$v_{\rm pec}(z=0) < H_0 \sqrt{\frac{2\Omega_0}{1+z_{\rm eq}}} R_0/{1
{\rm Mpc}}. $$
For $w=-1$, one obtains $v_{\rm pec}(z=0) < 47.1  R_0/{1
{\rm Mpc}}.$
For $w=-0.6$ one has $v_{\rm pec}(z=0) <42.8  R_0/{1
{\rm Mpc}}.$ This upper limit on the peculiar velocity is lower by
about 9\% in the quintessence model with $w=-0.6$ than in the
cosmological constant case. 

As pointed out by Baryshev et al. (2001), when $w<-1/3$ both linear modes
are decaying (the second, usually growing mode, decays as $a^{1+3w}$ 
for $z\ll z_{eq}$). Both density perturbations and peculiar velocities decay. 
The lower  the value of $w$ the faster is the decay, so  we would 
expect to have a colder Hubble flow in the $\Lambda$CDM model than in the QCDM
one, compensating for the earlier dominance of DE 
that we have in models with $w>-1$. These two contrary effects  
make hard to predict  which Dark Energy model will present a colder
hubble flow in a given environment.
The only way to test this scenario is via numerical simulations,
that we will present in the next section.

In cosmologies with no dark energy component like the open CDM model
(equivalent to $w=-1/3$), the zero-mass surface is pushed to infinity
and there are no decaying modes so peculiar velocities can't be
cooled, which is consistent with the high peculiar velocities seen in 
simulation of the OCDM model by Governato et al. (1997).

Finally, Axenides \& Perivolaropoulos (2002) have calculated the
growth suppression factor of velocity fluctuations in a dark energy
model compared to the Einstein-de Sitter model and concluded that the
presence of dark energy is not sufficient to explain the coldness of
the local flow.  However, this holds true for cosmological models with
different $w$ {\it and} the same high redshift  normalization of the density
field perturbations. However, realistic models  must  be normalized to match the
observed value of $\sigma_8$ at $z=0$.

From the above analysis it seems plausible that cosmologies with a dark
energy component would present colder flows around cosmic
haloes. However the magnitude of this effect, the dependence on $w$ and
on the local properties of the density field remain poorly constrained
by simple analytical arguments based on linear extrapolations.  These
considerations make resorting to numerical $N$--Body simulations
necessary.

\section{Simulations}
\label{s:iden}

We present  results from 
four simulations, performed using two different codes:

-- $\Lambda$CDM: two simulations, both with $h= 0.7$, $\Omega=0.3$, 
$\Omega_{\lambda}$=0.7, $\sigma_8=1.0$ 
but with different resolutions (see details below);

-- $\Lambda$WDM: a flat model similar to 
the low resolution $\Lambda$CDM 
but with a sterile neutrino mass of 2~keV as a warm dark matter candidate; 

-- QCDM: a quintessence model with the same parameter of the previous two simulations 
$\Lambda$C(W)DM model but with a constant
equation of state parameter for the dark energy $w = p_Q/\rho_Q=-0.6$. 

The $\Lambda$ simulations  were performed using 
{\tt PKDGRAV}, 
a parallel, multistepping tree-code with periodic boundary conditions 
(Stadel 2001). 
For the quintessence simulation, we used  
the parallel Adaptive Refinement Tree code 
{\tt ART} (courtesy of A. Klypin, Kravtsov, Klypin \& Khokhlov 1997). 

QCDM models with $w>-0.7$ are now disfavored by CMB observation (Kogut
et al. 2003, Verde et al. 2003). 
Nevertheless we have used this value for $w$ in order to try to maximize
the influence of dark energy on $\sigmah$ (so it will be used as a limiting case).
According to theoretical expectation (Baryshev et al. (2001) and
section \ref{sec:theo}) the strength of this effect depends on the
epoch when the DE becomes the dominant component of the universe: the
earlier the DE period begins, the stronger the cooling.  In our QCDM
model the DE period starts earlier than in the $\Lambda$CDM model, and
so we would expect a larger effect of dark energy (but see discussion
in earlier paragraph).  To calculate the power spectrum $P(k)$ for the
$\Lambda$WDM model and the QCDM used the {\tt CMBFAST} code to
generate transfer functions (Seljak \& Zaldarriaga, 1996).  Sterile
neutrinos are a popular candidate for warm dark matter (Dodelson \&
Widrow 1994, Dolgov \& Hansen 2002).  Though formation mechanisms for
sterile neutrinos are likely to produce a non-thermal energy spectrum,
the filtering effect may still be modeled accurately using thermal
neutrinos with adjusted masses (Hansen et al. 2001).  The $\Lambda$WDM
model can be characterized in terms of the reduction in fluctuations
on dwarf galaxy scales.  For 2~keV neutrinos the fluctuations on the
scale of $10^{11}$ M$_\odot$ ($\sigma_{0.59}$) are down by 10\% over
$\Lambda$CDM.  The power spectrum at $2 \pi/k = r = 0.59 h^{-1}$ Mpc
is lower by 2 orders of magnitude and thus contribution of small
objects to $\sigma_{0.59}$ is vastly reduced.  Fluctuations on the
scale of $10^{12}$ M$_\odot$ haloes are only marginally affected
($\sigma_{1.26141}$ down by 3\%).

Each  $\Lambda$C(W)DM simulation cube is 100~Mpc on a side with both using 144$^3$
particles and a spline kernel softening of 60~kpc (the spline kernel
is completely Newtonian at 2 softening lengths) 
The QCDM simulation has a box size of
90~Mpc with 128$^3$ particles, which gives roughly the same mass resolution
as the low resolution $\Lambda$ run.  Since our goal is
to identify galactic sized dark haloes, not to resolve their internal
structure, this combination of mass/force resolution is adequate.
Each simulation used at least 512 steps for every particle, and many
more (up to a few thousands) for particles with large accelerations
as those at the center of dense haloes.  The mass resolution is such
that a well resolved halo (30 particles) at redshift zero has a circular
velocity of 95~km/s. 
We can therefore resolve isolated haloes that are associated
with galaxies that have luminosities as low as a few \% of an $L_\star$
galaxy (Cole et al. 2001).  

We have also analysed a very high resolution simulation for the
$\Lambda$CDM model, with $81\times 10^6 (432^3)$ dark matter particles
simulated from a starting redshift $z_0=69$. The employed volume is
50~$h^{-1} $Mpc, which gives a particle mass of $1.3 \times 10^8
h^{-1}$ M$_{\odot}$ allowing us to resolve haloes down to less than
$10^{10} h^{-1}$M$_{\odot}$ with 75 particles. The force resolution is
5~$h^{-1}$ kpc and the particles in the highest density region undergo
20~000 timesteps (this simulation is described in more details in Reed
et al. 2003a). In this run, we are able to resolve the substructure
inside haloes corresponding to galaxy groups, allowing us to evaluate
the effect of substructure within larger haloes in velocity dispersion
measurements.  The results of a comparison with the low resolution run
are shown in Fig.~\ref{fig:resol}.

Our $\Lambda$CDM model has parameters very close to those currently
suggested by observational results.  This model has been successful on
an extended number of crucial tests on scales down to a few Mpc.
However, recent results pointed out a possible flaw of $\Lambda$CDM at
galactic and subgalactic scales, like steeper density profiles than
those implied by galaxy rotation curves (Moore et al. 1998, de Block
\& Bosma 2002, Swaters et al. 2003, Reed et al. 2003b, Diemand et
al. 2004a) and an excess of substructure (i.e. small satellites) in
galactic haloes (Moore et al. 1999, Klypin et al. 1999, Ghigna et
al. 2000, De Lucia et al. 2004, Diemand et al. 2004b; Klypin et al.
(2003b) for QCDM model).  Both problems may be linked to an excess of
power at small scales.  Hence, as a cure to the above mentioned
problem a WDM rather than CDM component reduces the amount of density
fluctuation and subgalactic scale and may prove useful in cooling the
Hubble flow around the Local Group.  To boost its possible effects we
used for the mass of the WDM a value of 2~keV, this is close to the
minimum allowed from constraints based on satellites abundance and the
abundance of absorption systems (Bode, Ostriker \& Turok 2000). A
lower WDM mass would likely underproduce both of them. Recently
Yoshida et al. (2003) put strong constraints on WDM candidates lighter
than 10keV based on the abundance of haloes at the epoch of
reionization.

\section{Halo finders and selection of LG systems}

Our simulations only follow the evolution of the dark matter
component. It is expected that baryons condense and form galaxies at
the centers of the dark matter haloes (White \& Rees 1978; Cen \&
Ostriker 1992,1993a,b; Katz, Hernquist \& Weinberg 1992).  For our
purpose of locating LG halo candidates (not resolving their internal
structure), we resorted to two algorithms: 
{\tt SO} and {\tt SKID} ({\tt SKID} is publicly available at 
http://www-hpcc.astro.washington.edu).
Our version of {\tt SO} (Lacey \& Cole 1995) uses {\tt SKID} halo centers (based on the
density maxima) as a start, then a sphere is grown around this center until
the enclosed density drops below some threshold which is chosen to be
that associated with the average density inside a spherical virialised
halo in that given cosmology (values were $97.1 \rho_{\rm crit}$ and $135.2
\rho_{\rm crit}$ for LC(W)DM and QCDM respectively (Mainini et al. 2003); 
At difference with {\tt SKID}, {\tt SO} imposes a spherical symmetry to the haloes 
and, perhaps more importantly, regroups substructure within larger virialized structures.
We verify that this  does not affect our results in subsection 6.2.
We verified that our results were largely independent on which halo finder
we used and decided to show results only from {\tt SO}.  In this paper we
assume that a galaxy with a certain circular velocity
would be found in a dark halo that has a similar circular velocity (Cole et al. 2001).

\section{STATISTICAL PROPERTIES OF LOCAL GROUP CANDIDATES: COMPARING 
simulations with  THEORY and 
  OBSERVATIONS}

\subsection{Selection of Local Groups}
Following G97, from the halo catalogs obtained we selected ``Local Groups'',
i.e. binary haloes with the following criteria:

\begin{enumerate}
\item{ A generic sample of binary haloes with separations $s <
1.5$ Mpc and circular velocities $125 < V_c < 270$ km/sec.}
\item A LG sample defined such that $s < 1.0$ Mpc,
circular velocities $125 < V_c < 270$ km/sec, negative radial
velocities and no neighbors within 3 Mpc with circular velocity larger
than either of the two group members.
\item  A LG sample defined as (2) but 
with the additional requirement that the binary haloes must lie close 
(5--12$h^{-1}$Mpc) to a Virgo sized cluster 
with $500 < V_c< 1500$~km/sec.
\end{enumerate}    

In the $\Lambda$CDM cosmology, 
about 30 Local Groups satisfy criteria 1) and 2), 
and about 15 satisfy point 3) as well. 
As we will show later there are no significant differences between samples 2 and 3
(and actually with the global field halo population, see Fig.~\ref{fig:all}), 
and we will refer to our Local Group sample as systems identified using
criterion 2).

We define the peculiar line-of-sight velocity dispersion from the Hubble flow
for a set of haloes within a sphere of radius $R$ (in Mpc) centered on the center
of mass of each simulated LG as:
\begin{equation}
\sigmah = \sqrt{
\frac{1}{n-1}
\sum_{i=1}^n
{\left[ \left( v_i - H_{\rm loc} D_i \right) - \langle {v-H_{\rm loc}D}
    \rangle \right]}^2} 
\label{eq:HF}
\end{equation}
where 
\begin{equation}
\langle{v-H_{\rm loc}D}\rangle = \frac{1}{n} \sum_{i=1}^n (v_i - H_{\rm loc}
D_i)
\end{equation}
where a local Hubble constant H$_{\rm loc}$ is fit to each dataset of haloes.
We have used the same procedure also to analyse the data points shown in 
Fig.~\ref{fig:rho}.

\subsection{Comparison with observations}
\begin{figure}
\centering
\epsfxsize=\hsize\epsffile{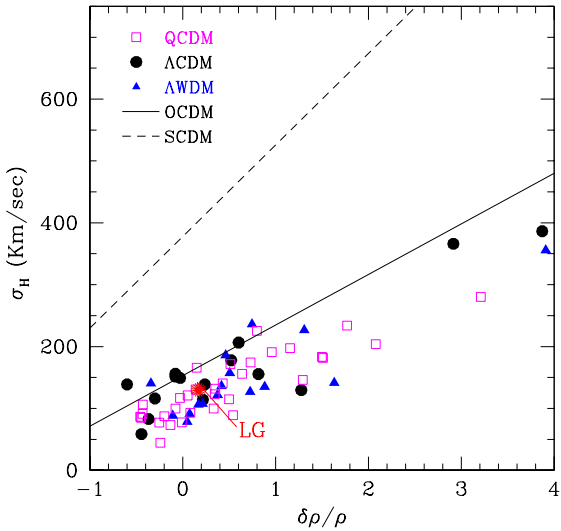}
\caption{
Hubble flow velocity dispersion in a sphere of 10~Mpc: $\Lambda$CDM (circles), $\Lambda$WDM 
(triangles) and QCDM (squares). 
The lines marks the results from models described in G97, the large starred
dot marks the real Local Group.}
\label{fig:overd}
\end{figure}

Comparing the results from this new set of simulations to the ones
obtained by G97 (see Fig.~\ref{fig:overd}), it is evident that the
new, dark energy dominated models fare better, expecially when
compared to the new observational estimate of $\sigmah$.  The velocity
dispersion is down by at least a factor of two compared to the
classical, but outdated SCDM model for a large range of overdensities.

\begin{figure}
\centering
\epsfxsize=\hsize\epsffile{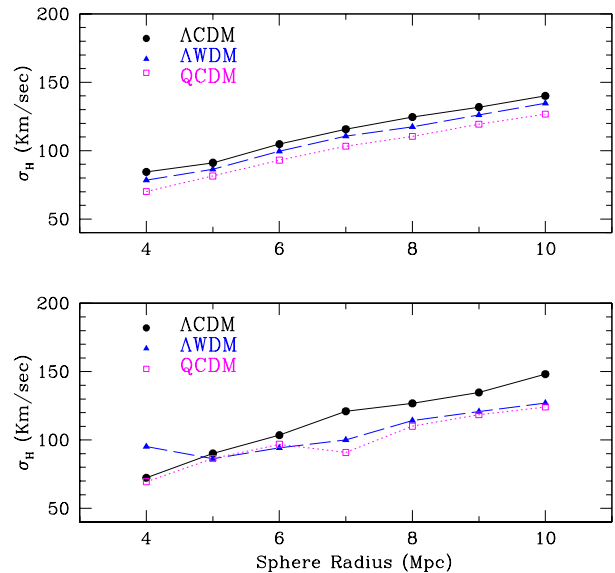}
\caption{
Hubble flow velocity dispersion  
computed in spheres of growing radius for different cosmological
models. The lower panel shows results for LGs with $-0.1<\delta\rho/\rho< 0.6$, 
the upper panel for all the field galaxy-sized haloes.}
\label{fig:cfr}
\end{figure}

$\Lambda$CDM and $\Lambda$WDM give almost the same $\sigmah$ for the whole 
range of overdensities. The QCDM model  produces a colder
Hubble flow, as better shown in Fig.~\ref{fig:cfr}.

\begin{figure}
\centering
\epsfxsize=\hsize\epsffile{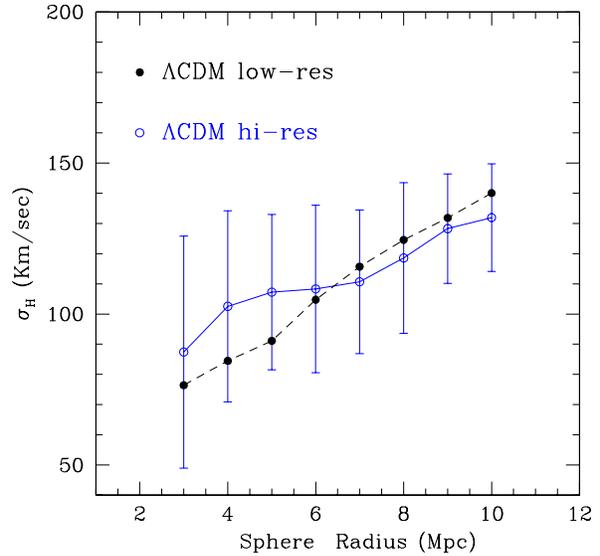}
\caption{
Hubble flow velocity dispersion  
computed in spheres of growing radius for the  
high-resolution $\Lambda$CDM simulation.}
\label{fig:resol}
\end{figure}

In Fig.~\ref{fig:resol} we show the comparison between low and high resolution
simulations. 
As expected, the high resolution run has a higher value of $\sigmah$ at
small scales, due to the fact that we are able to resolve the
inner structure of the more massive haloes (corresponding to galaxy groups)  around the LGs. 
The velocity dispersion is lower on scales $> 7$ Mpc. 
This could be
related to the smaller box size that reduces the effect of the
large-scale structure. Nevertheless, both simulations agree  
within the 1$\sigma$ error bars showing that the effect of substructure is small.

As shown by G97, the velocity dispersion around the Hubble flow depends
strongly on the local overdensity. This can be determined for the Local
Group using the complete catalogue of IRAS galaxies.  The current best
estimate based on IRAS galaxies is $\delta \rho /\rho=0.60\pm 0.15$ for a
top-hat sphere of radius 500~km/s (Strauss 1996, private communication).
Hudson (1993) uses a compilation of optical galaxy surveys to study the
local density field and within the same volume he finds an overdensity of
$\sim 0.2$.  The agreement between the optical and IRAS surveys is
encouraging and indicates that the Local Group resides in a region of the Universe of
almost average density.

\begin{figure}
\centering
\epsfxsize=\hsize\epsffile{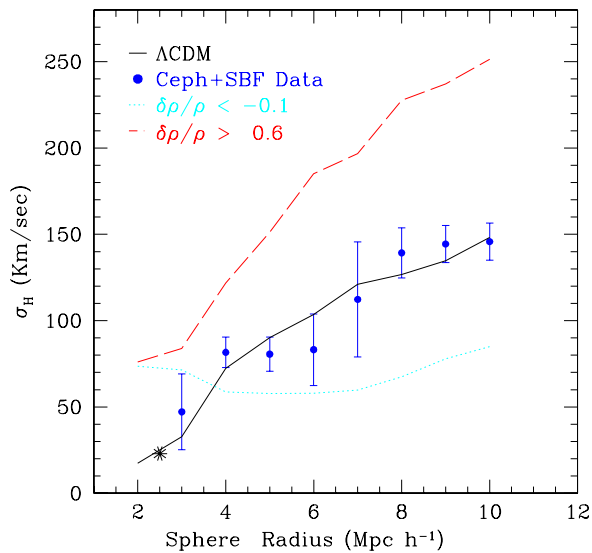}
\caption{Hubble flow velocity dispersion computed in spheres of growing radius. 
The large filled dots mark results from our new estimate of $\sigmah$ for
the Local Group, the last starred point was taken from the estimate by  
Karachentsev et al. (2002). Simulation results are shown for the $\Lambda$CDM
model for different choices of the density environment.
Note that the flattening of the dotted curve at small $R$s is due to the low number of 
LGs in that overdensity range. The first two points (R=2 and R=3) are made both
with just 2 LGs, so the statistical variance is quite high.
}
\label{fig:rho}
\end{figure}

If simulated LGs are selected in regions with a local overdensity between 
$-0.1<\delta\rho/\rho< 0.6$, both the normalization and the shape
of the $\sigmah$ vs $R$ relation are well reproduced (see Fig.~\ref{fig:rho}) 
-- a remarkable success for flat, $\Lambda$-dominated models.

The $\Lambda$WDM performs equally well (cf Fig.~\ref{fig:cfr}, the higher value of $\sigmah$
at very small scales is of small statistical significance), 
indicating that it is the presence of a cosmological constant, 
which is the most important ingredient, rather than the details of the power 
spectrum at subgalactic scales. This is further supported by the QCDM simulation, 
which produces a Hubble flow colder than in both $\Lambda$ models at all scales.
That seems to indicate that a longer period of dark energy dominance has the capability 
to produce a colder Hubble flow, in agreement with the theoretical expectations 
by Baryshev et al. (2001) and section \ref{sec:theo}).
Comparing our analysis to the results of G97, it appears that 
only DE-dominated
models are able to satisfy at the same time small and large scale
structure constraints {\it and} generate a substantial number of simulated LGs
with $\sigmah$ comparable to the observed values. 
Also, LGs in LC(W)DM cosmologies fail to have $\sigmah$ in the correct range
if their local overdensity is too high or too low, as shown in Fig.~\ref{fig:rho}.

How are important the details of the mass distribution around the LG
other than the average density on a few Mpc scale? Klypin et al. 
(2003a) obtained a cold Hubble flow around a simulated region that
replicated the mass distribution inferred from a redshift survey of
the local Universe. Their simulation assumed a $\Lambda$CDM cosmology. 
Their work was based on one constrained simulation of our local environment,
our LG samples were selected using some simple
selection environment criteria like the absence of nearby massive
groups or the presence of a rich cluster within 5--12~Mpc. However, no
substantial difference was found between the typical $\sigmah$ in the
different subsamples depending on the presence of nearby clusters.

\begin{figure}
\centering
\epsfxsize=\hsize\epsffile{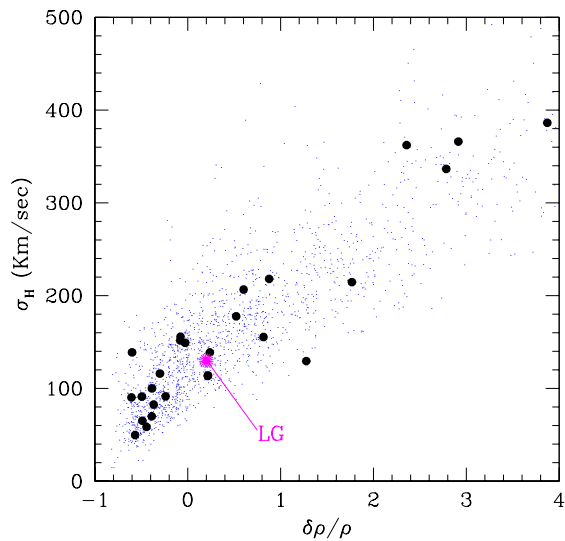}
\caption{Peculiar velocity around LG candidates 
and field galaxy-sized haloes (small dots) in the $\Lambda$CDM model }
\label{fig:all}
\end{figure}

To further test the hypothesis that a special local mass distribution
is necessary for a cold Hubble flow, we repeated our analysis for a
much larger sample of Milky Way (MW) sized field haloes, with no
special selection criteria. Results are shown in Fig.~\ref{fig:all}. A
generic sample of MW-sized field haloes shows values of $\sigmah$ in
the correct range if the local environment density is around
average. This shows that the Local Group environment and details of
its mass distribution are not special in any significant way and that
in DE-dominated
cosmologies a small $\sigmah$ is a general
characteristic of the environment around field galaxies.

\section{Conclusions}

Observational evidence at large and intermediate scale points to a
flat, $\Lambda$-dominated CDM Universe.  The coldness of the Hubble
flow provides another, independent  test at intermediate scales, where different
kinds of dark matter and energy could play a significant role in
shaping the local dynamics.

In this paper we have tested the predictions from cosmological models
now strongly supported by observational evidence against a new, robust
estimate of the Local Group $\sigmah$ in spheres of increasing radius
up to 10~Mpc based on a sample of nearby galaxies.

We have shown that simulations are able to successfully
reproduce the observed cold Hubble flow around field galaxies only in a flat,
DE ($w \le -1$)
dominated cosmology and only in environments where the local density
measured on a scale of a few Mpc is close to  the average one.
The inclusion of a warm dark matter component does not change
this result significantly, as removal of small-scale power does not
play a major role.

We have also shown that $w>-1$ (quintessence models) where the era
dominated by dark energy  begins earlier, produce  an Hubble
flow marginally colder than  $\Lambda$CDM.

This is in agreement with the theoretical expectation (see section
\ref{sec:theo} and also Baryshev et al. 2001, Chernin et al. 2004)
that the coldness of the local Hubble flow follows from the fact that
a galaxy's peculiar velocity decreases via adiabatic cooling if it
spends a sufficient time in the dark energy dominated region where
linear perturbations decay.  The presence of a dark energy field is
then the key ingredient to explain the observed coldness of the local
Hubble flow.

While the right cosmology and environmental density are crucial, the
local dynamical criteria do not single out the LG from the generic
population of field galaxies. A cold Hubble flow around Milky Way
sized galaxies is equally common as in our LG subsample.  This implies
that the details of the mass distribution in our local environment,
like the vicinity to larger clusters and/or voids and filamentary
structures, do not play a major role in shaping its velocity field.

The environment of our Local Group provides a new, independent
evidence for the existence of dark energy on scales of just few Mpcs, 
corroborating the evidence gathered from observations of distant
objects and the early Universe.

\section*{Acknowledgments}
 
The authors thank Jeff Gardner for the use of his SO halo finder and Anatoly
Klypin for the use of the ART code for the QCDM run. We also thank Tom Quinn,
Ben Moore, Joachim Stadel, James Wadsley and Vandana Desai for useful
discussions. FG is Brooks Fellow. FG was supported in part by NSF grant AST-0098557. 
Computations were carried out at CINECA and ARSC supercomputing centers.
CH acknowledges financial support from the Swedish Research Council
{\it Vetenskapsr\aa det}.

\end{document}